\documentclass[aps,prb,twocolumn,superscriptaddress]{revtex4-2}

\usepackage{graphicx}
\usepackage{dcolumn}
\usepackage{bm}

\usepackage{graphicx}
\usepackage{derivative}
\usepackage{dcolumn}
\usepackage{orcidlink}
\usepackage{bm}
\usepackage{hyperref}
\usepackage{braket}
\usepackage{amsmath}
\usepackage{xcolor}

\usepackage{amsfonts}
\usepackage[dvipsnames]{xcolor}

\begin{document}

\preprint{APS/123-QED}

\title{Exact mobility edges in a slowly varying quasiperiodic ladder model}

\author{Arpita Goswami\,\orcidlink{0009-0003-2241-7034}}
\affiliation{Department of Physics, Indian Institute of Technology Tirupati, India, 517619}

\date{\today}

\begin{abstract}
We propose a minimal two-leg ladder model in which the mobility edge (ME) arises solely from bond modulation, induced by a slowly varying quasiperiodic modulation of the inter-leg tunneling amplitudes. We demonstrate that this bond-modulated ladder naturally hosts two propagation channels, whose symmetric and antisymmetric combinations experience opposite effective onsite potentials. Using the adiabatic (slowly varying) limit of the modulation, we derive an exact analytical condition for the single-particle mobility edge, $E_c=\pm|2t-\lambda|,$ where $t$ is the hopping amplitude along both the legs and $\lambda$ is the bond modulation strength. This result directly generalizes the classic ME condition for slowly varying onsite potentials to a multi-leg (two-leg in our case) geometry. Extensive numerical calculations, including inverse participation ratios, Lyapunov exponents, density of states, and participation-ratio scaling, demonstrate excellent agreement with the analytical prediction across a wide range of parameters. We further identify a regime for small modulation exponents $0 < \nu <1$, where localized and weakly delocalized states coexist even beyond the transition point ($\lambda_c=2t$). Finally, we characterize the crossover from the slowly-varying regime to a true quasiperiodic regime where the adiabatic approximation breaks down, supported by a thorough scaling analysis of the generalized participation ratio and the multifractal properties of the eigenstates. Our results establish that a deterministic bond modulation can serve as a sufficient ingredient to produce an exact ME in ladder systems, offering experimentally accessible and advantageous routes toward tuning nonergodic extended phases.

\end{abstract}


\maketitle

\section{Introduction}

The interplay between order, quasiperiodicity, and disorder \cite{anderson.1958} is a central theme in condensed matter physics. Quasiperiodic lattices serve as a unique testbed between perfectly periodic crystals and truly disordered solids, and therefore display a distinct nature of localization phenomena that are absent in either limit. Models with correlated deterministic modulations including quasiperiodic disorder~\cite{quasi_1, quasi_2, quasi_3, quasi_4, quasi_5, quasi_6, quasi_7,aa_addi3, HB,quasi_rev_1} exemplify this intermediate regime. A paradigmatic example is the Aubry-Andr\'e-Harper (AAH) model~\cite{Aubrey_1}, which exhibits a sharp localization transition at a finite critical strength of the onsite quasiperiodic potential ($\lambda=2t$, with $\lambda$ the potential strength and $t$ the nearest-neighbor hopping). However, due to its exact self-duality, no ME (where localized and delocalized states coexist for a particular potential strength) can occur in the spectrum: all eigenstates are either fully extended or fully localized depending solely on $\lambda$, and at the critical point, the entire spectrum becomes multifractal. Quasiperiodic systems can also host a critical phase characterized by eigenstates that are spatially extended yet nonergodic, exhibiting scale-invariant (multifractal) structure. Such states are distinguished by nontrivial scaling of participation ratios, $P_2 \sim L^{-D_2}$ with $0 < D_2 < 1$, in contrast to fully extended ($D_2 = 1$) or localized ($D_2 = 0$) states. This leads to anomalous spectral statistics, fractal wavefunction structure, and subdiffusive dynamics~\cite{critical_1,critical_2,critical_3,critical_4}. In the presence of interactions, this regime can give rise to a many-body critical phase, often referred to as a nonergodic metallic phase, which lies between the thermal and many-body localized phases~\cite{nem_modak,nem_soumi,deng2017many,nandkishore2015many}. In this work, we use the term ``nonergodic metallic-like phase'' to denote the coexistence of localized and extended states in energy, where the extended states themselves exhibit multifractal scaling and thus remain nonergodic.

Achieving a single-particle mobility edge (SPME), i.e., energy-dependent coexistence of localized and extended states, thus requires breaking this self-duality. Several routes have been developed, including long-range hopping~\cite{long_range_hopp_exp}, slowly varying or nonlinear onsite modulations~\cite{duality_breaking, sv_sds_1, sv_sds_2, int_sv}, and non-Hermitian extensions~\cite{NH_exact_ME_4, PT_exact_ME_3}. More recently, duality concepts have been generalized to complex deterministic modulations, enabling analytically predictable MEs in models with generalized AA potentials \cite{NNN_AA}, mosaic lattices \cite{Mosaic_exact_ME_2, Mosaic_ME_6}, and quasiperiodic networks~\cite{Exact_ME_1, exact_ME_5}. In these cases, breaking self-duality is typically achieved through modified onsite potentials, exponential-couplings~\cite{long_range_hopp_exp}, multiple incommensurate frequencies~\cite{gen_AA}, or power-law tunneling~\cite{power_hop_ME_7, Long_range_AAH, Shallow_QP}.

Within this class of deterministic modulations, slowly varying onsite potentials provide a particularly clean route toward analytic MEs. Seminal works by Das~Sarma \textit{et al.} demonstrated that their adiabatic, locally constant character yields transparent analytical prediction of MEs~\cite{sv_sds_1, sv_sds_2}. In one dimension, this concept has recently been extended to off-diagonal modulations \cite{off_1d}, and network-like structures~\cite{Exact_ME_1, long_range_kitaev_AAH, exact_ME_5,hidden_duality}. A previous work on the quasiperiodic ladder also introduced quasi-periodic modulation and demonstrated ME; however, the origin of the ME stems from the distinct hopping of the two legs \cite{coupl_chain}. In particular, Ref.~\cite{coupl_chain} studied coupled Aubry–André chains in which each leg carries onsite quasiperiodicity. By contrast, in our model, both legs remain completely clean, with no onsite quasiperiodicity on either leg.

Existing other quasiperiodic ladder studies rely on onsite quasiperiodicity on at least one leg \cite{coupled_ref_3} or explicitly introduce disorder asymmetry between the channels \cite{coupled_1d_sds,coupled_1d_ref_2}. In such models, MEs originate from a modified onsite energy landscape.

Despite these advances, an important question has remained unanswered: Can SPMEs arise in a minimal ladder system purely from quasiperiodicity in inter-leg hopping without any onsite modulation, disorder, or long-range hopping on either leg?

In this study, we address this issue by considering a ladder model in which both legs remain completely clean, with the only source of quasiperiodicity arising from deterministic modulation of the rung hoppings (this ladder construction we define as the slowly varying quasiperiodic ladder (SVQL) model). In the slowly varying regime ($0 < \nu < 1$), the ladder admits an analytical treatment within the locally constant approximation, yielding the exact mobility-edge condition
\[
E_c=\pm|2t-\lambda|.
\]
 The present work demonstrates that exact mobility edges can be engineered using purely bond-modulated couplings without introducing explicit onsite quasipericity on either leg. Importantly, the ladder geometry plays an essential role in the present construction. In contrast to a single one-dimensional chain, where quasiperiodic modulation of hopping amplitudes alone does not generally produce the mobility-edge structure obtained here, the two-leg ladder converts the rung modulation into effective quasiperiodic channels through the symmetric–antisymmetric decomposition, the effective channel Hamiltonian coincides with forms previously analyzed in Ref.~\cite{sv_sds_1}, yet the microscopic ladder construction remains distinct. The resulting mobility edges, therefore, arise from the interplay between bond modulation and the multi-channel (two-channel in this study) ladder geometry. We further characterize the crossover from the slowly varying regime to the Aubry–André limit by analyzing inverse participation ratios. We also present a detailed multifractal scaling analysis over a wide range of parameters for the ladder geometry. This study further presents a bond-modulated ladder realization that exhibits the coexistence of localized and multifractal extended states over a broad parameter regime. In an interacting setting, such coexistence may provide a useful platform for exploring nonergodic metallic behavior, where states can exhibit volume-law entanglement growth while failing to thermalize~\cite{ETH1, ETH2, rigol2012alternatives}(called fractal in the non-interacting limit). This highlights experimentally accessible pathways for controlling the coexistence of localized and extended states in low-dimensional lattices.
We verify our results numerically using a comprehensive set of diagnostics, inverse participation ratios (IPR), Lyapunov exponents, singular features in the density of states, and participation-ratio scaling. All methods consistently reveal coexisting localized and extended states in energy, forming a nonergodic metallic-like phase~\cite{metal_1, metal_2, metal_3, metal_4}. The sharpness of the MEs improves as $\nu$ decreases, reflecting enhanced adiabaticity of the modulation.

The remainder of this paper is organized as follows. Section~\ref{sec2} introduces the model and its ME heat-map. Section~\ref{sec3} presents the analytical derivation of the ME condition. Section~\ref{sec4} contains numerical verification. In Sec.~\ref{sec_rev_1} we identify the region where the locally constant approximation breaks for a finite-size system. Section.~\ref{sec_rev_2} we present a detailed multifractal analysis of the system in different parameter regions. Section~\ref{cons} discusses experimental relevance and possible extensions to multiband quasiperiodic systems.


\begin{figure}[h!]
\includegraphics[width=1\linewidth]{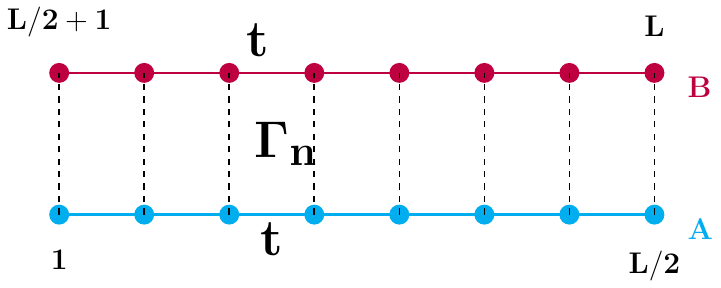}
    \caption{Schematic representation of the Slowly Varying Quasiperiodic Ladder (SVQL) model. The system consists of two parallel tight-binding chains (legs), coupled through inter-leg rung hopping amplitudes $\Gamma_n= \lambda \cos{(2Qn^{\nu})}$, which incorporate a deterministic slowly varying quasiperiodic modulation. Horizontal links represent uniform intra-leg hopping t, which is the same in both legs.}
    \label{fig-1}
\end{figure}
\section{Model and A heatmap of average IPR}
\label{sec2}
The schematic representation of the ladder system is shown in Fig.~\ref{fig-1}. The model consists of two parallel tight-binding chains (legs A and B), coupled through vertically aligned rung hoppings. There are two types of tunnelling processes: (i) nearest-neighbor hopping $t$ along each leg between the $n^{\mathrm{th}}$ and $(n+1)^{\mathrm{th}}$ sites, and (ii) inter-leg hopping between the two legs at site $n$, modulated by a slowly varying quasiperiodic function. The full Hamiltonian reads
\begin{equation}
\label{ham_eq}
H = t \sum_{n}\left(A^{\dagger}_{n+1}A_{n}+B^{\dagger}_{n+1}B_{n}\right) + \mathrm{H.c.}
+ \Delta_{1},
\end{equation}
with
\begin{equation}
\Delta_{1} = \sum_{n} \Gamma_{n}\left(A^{\dagger}_{n}B_{n} + B^{\dagger}_{n}A_{n}\right).
\end{equation}

The rung modulation $\Gamma_n$ is neither random nor strictly periodic but deterministic and quasiperiodic in nature. It takes the form
\begin{equation}
\Gamma_{n} = \lambda \cos(2Q\, n^{\nu}),
\end{equation}
where $\lambda$ controls the contrast in the inter-leg coupling, $Q$ is a real number, and $0<\nu<1$ denotes the slowly varying exponent. Throughout this work, we set $Q = (\sqrt{5}-1)/2$, such that in the limit $\nu \rightarrow 1$ the model continuously connects to the modulation used in Harper’s equation~\cite{harper_eq}. For numerical simulations, we vary $\lambda$ within the range $0 \leq \lambda \leq 10$ and typically choose the total system size $L = 2000$, ensuring convergence to the thermodynamic limit. We use the exact diagonalization technique for the numerical part of the study and use system sizes up to $L=20000$ to confirm the stability of ME.

\begin{figure}[t]
    \centering
    \includegraphics[width=0.98\linewidth]{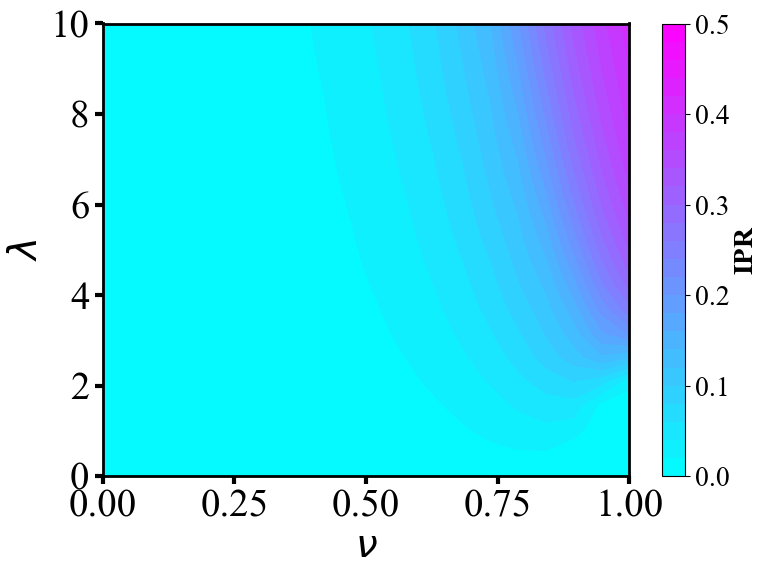}
    \caption{A heatmap of average IPR for the SVQL model in the ($\nu-\lambda$) parameter space. The diagram displays the emergence of distinct spectral regimes: extended (blue) and localized (red). }
    \label{fig:phase_diag}
\end{figure}

To characterize the spectral phases of the system in the $(\nu,\lambda)$ parameter space, we compute the average inverse participation ratio (IPR), as shown in Fig.~\ref{fig:phase_diag} (details of IPR calculation are provided in Sec.~\ref{sec4}). The exponent $\nu$ determines the degree of adiabaticity in the quasiperiodic modulation, while $\lambda$ tunes the strength of the two-channel hybridization. For small $\lambda$ and $\nu$, most states are extended. Increasing either parameter (beyond $\lambda_c$) drives the system into a localized regime due to enhanced channel interference. Between these two limits, a broad intermediate region emerges in which localized and extended eigenstates coexist in energy, signalling the presence of a ME-induced nonergodic metallic-like phase. Notably, the analytically derived ME condition (Sec.~\ref{sec3}) matches the numerically obtained phase boundaries in \ref{sec4} with excellent accuracy. This heat map provides a qualitative understanding of the system's spectral nature in the $\nu-\lambda$ parametric region. We conduct an extensive analytical and numerical study to gain a more detailed understanding of the ME locations. The details of the channel formation have been elaborated in Appendix \ref{app_chan}.

\section{Predicting metal-insulator transition in the system through analytical calculation} \label{sec3}

\begin{figure*}[t]
    \centering
\includegraphics[width=1\linewidth]{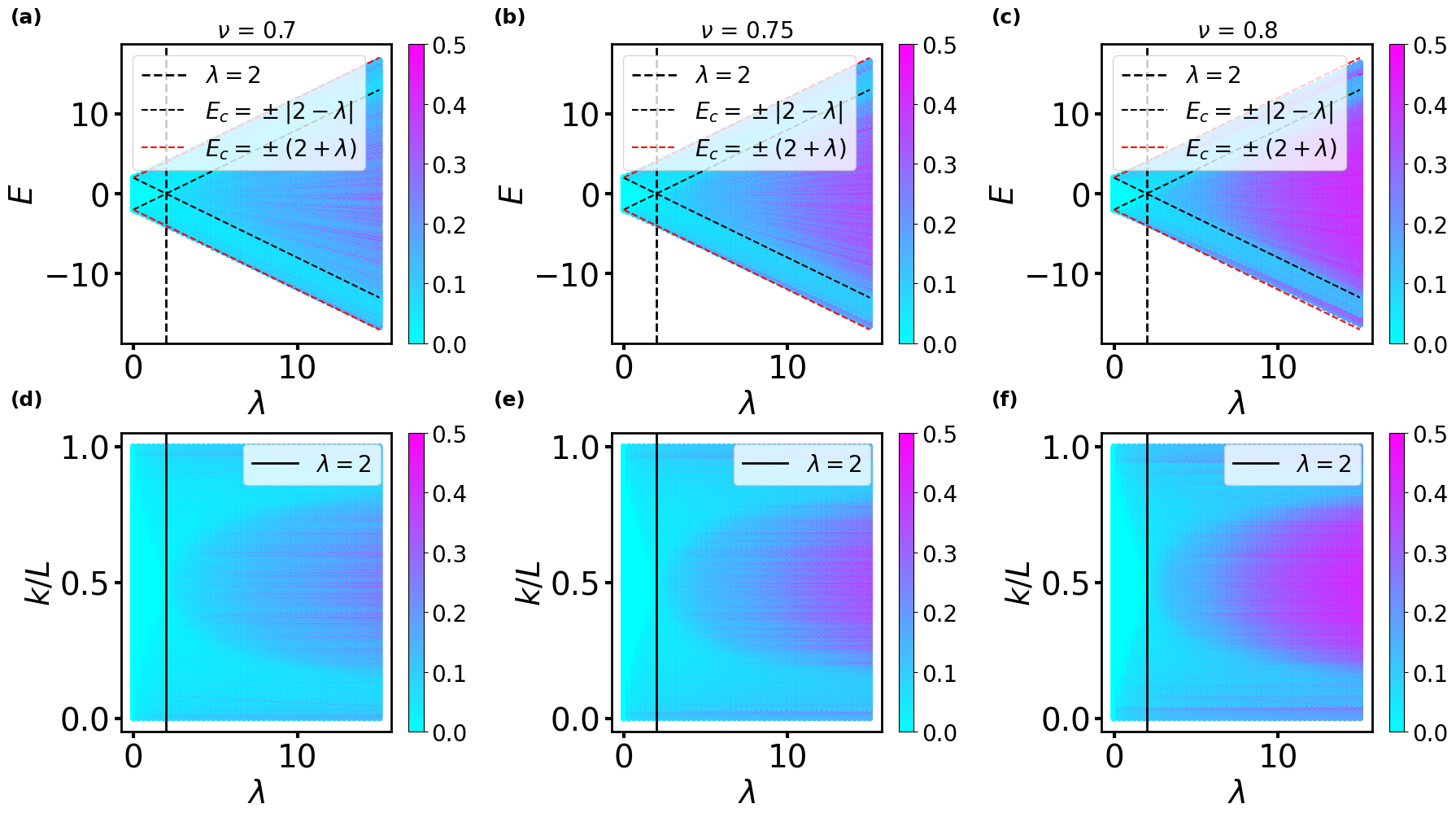}
    \caption{(a)-(c) are the contour plots of $IPR$ as function of $\lambda$ and energy($E$) and (d)-(f) are the contour plots of $IPR$ as function of $\lambda$ and state index ratio($\frac{k}{L}$, where $k$ is the state index and $L$ is the total number of site) for all eigenstates. Calculations are performed for the total system size (L) = 2000 and for $\nu$ = 0.7, 0.75, and 0.8, respectively; $t$ is fixed to 1 for the calculations. The plots reveal clear energy-dependent MEs whose position is $\nu$ independent, confirming that slower quasiperiodicity enhances localization contrast and sharpens the extended-localized spectral separation.}
    \label{fig:IPR_lambda}
\end{figure*}
Due to its highly correlated nature, the slowly varying quasiperiodic modulation brings a new perspective to 1D systems. Moreover, to understand the localization phenomena in them, the researchers developed an asymptotic semiclassical Wentzel-Kramers-Brillouin (WKB) technique to find the density of states and Lyapunov exponent, and they observed that these two quantities are not smooth or have abrupt jumps across the MEs, which is absent in the 3D Anderson transition \cite{anderson_3d}. Using semiclassical approximations, we derive the exact expressions for MEs and verify them using several numerical results.

 The Hamiltonian in Eq.~\ref{ham_eq} takes the following one-particle form in operator notation:
\begin{equation*}
\begin{split}
    \hat{H}  = & \, t \sum_{n}(\ket{A_{n+1}}\bra{A_{n}}+\ket{A_{n}}\bra{A_{n+1}}\\
    &+ \sum_{n}\ket{B_{n+1}}\bra{B_{n}}+\ket{B_{n}} \bra{B_{n+1}}) \\ 
    & +\sum_{n}\lambda \cos{(2Q n^{\nu})}(\ket{A_{n}} \bra{B_{n}}+\ket{B_{n}} \bra{A_{n}}).
\end{split}
\end{equation*}  

We expand the wavefunction in terms of the site basis of the two legs as $ \ket{\psi}= \sum_n a_n \ket{A_n} + b_n \ket{B_n}$, where $\ket{A_n}$ and $\ket{B_n}$ represent sites on legs A and B, respectively, and $a_n$, $b_n$ are the corresponding amplitudes. Evaluating $\bra{A_j}\hat{H}\ket{\psi}$ and $\bra{B_j}\hat{H}\ket{\psi}$ leads to the following coupled equations for the coefficients:
\begin{equation} \label{eq1}
t a_{j+1}+t a_{j-1}+\Gamma_j b_j =E a_j,
\end{equation}
\begin{equation} \label{b_coeff}
t b_{j+1}+t b_{j-1}+\Gamma_j a_j =E b_j,
\end{equation} 
where $\Gamma_j =\lambda \cos{( 2 Q j^{\nu})}$. These equations couple the amplitudes on the two legs. To proceed, our aim is to decouple them by using the following recurrence relations derived from Eq.~\ref{eq1}: 
\[
    t a_{j+1}+t a_{j-1}+\Gamma_j b_j =E a_j,
\]
\[
    t a_{j}+t a_{j-2}+\Gamma_{j-1} b_{j-1} =E a_{j-1},
\]
\[
    t a_{j+2}+t a_{j}+\Gamma_{j+1} b_{j+1} =E a_{j+1}.
\]

In the regime $0 < \nu < 1$, the modulation $\Gamma_j$ varies slowly and can be regarded as locally constant for thermodynamically large system size. This can be seen explicitly from the following:
\begin{equation}
\Gamma_{j} = \lambda \cos(2Q j^{\nu}),
\end{equation}
taking the derivation w.r.t $j$ gives the following form:
\begin{equation}
\frac{\partial \Gamma_{j}}{\partial j} = -2\nu\lambda  Q j^{\nu -1} \sin(2  Q j^{\nu}),
\end{equation}
so that
\[ \bigg |\frac{\partial \Gamma_{j}}{\partial j}\bigg| = \frac{ \nu\lambda Q  |\sin( Q j^{\nu})|}{j^{1- \nu}}.\]
As $j \to \infty$ with $0<\nu<1$, the derivative vanishes. Not only that, for $j \to \infty,$ the difference between $\Gamma_j$ and $\Gamma_{j+1}$ almost vanishes. Therefore, $\Gamma_j$ may be approximated as a locally constant term. A similar "local constancy" in the onsite potential concept is used in Ref. \cite{sv_sds_1}, which shows that the argument can systematically predict the localization transition for a large system. Hence, we set
\[
\Gamma_{j-1} \approx \Gamma_j \approx \Gamma_{j+1} \approx C.
\]

Substituting this approximation into Eq.~\ref{b_coeff} and eliminating $b_{j-1}$, $b_j$ and $b_{j+1}$ through the recurrence relation yields the following:
\begin{equation} \label{final_a_coeff_eq}
    t^2 a_{j+2}-2Et a_{j+1} +(E^2 +2t^2-C^2)a_j -2Eta_{j-1}+t^2 a_{j-2}=0.
\end{equation}
Dividing by $t^2$ and assuming a trial solution of the form $a_j \sim z^j$, we obtain the following:
\begin{equation} \label{comb_1}
    z^4 -2E^{'} z^3 +({E^{'}}^2 + 2 - {C^{'}}^2)z^2 -2 E^{'}z + 1 = 0,
\end{equation} 
where $E/t =E^{'}$ and $C/t = C^{'}$. It is to be noted that the assumption of the trial solution of the above form is valid in the locally constant approximation where the recurrence coefficients are site-independent. Dividing Eq.~\ref{comb_1} by $z^2$ gives as follows:
\begin{equation} \label{quart_quard} 
z^2 -2E^{'} z +({E^{'}}^2 + 2 - {C^{'}}^2) -\frac{2 E^{'}}{z} + \frac{1}{z^2}  = 0.
\end{equation}
\[ z^2 + \frac{1}{z^2} -2E^{'}( z +\frac{1}{z})+({E^{'}}^2 + 2 - {C^{'}}^2)  = 0\]
At this moment, the equation looks complicated and a non-trivial polynomial to solve.
However, the substitution $m = z + \tfrac{1}{z}$ (so that $z^2 + \tfrac{1}{z^2} = m^2 -2$) can reduce Eq.~\ref{quart_quard} to the following quadratic form
\[
m^2 -2 - 2 E^{'}m +({E^{'}}^2 + 2 - {C^{'}}^2)  = 0,
\]
or equivalently
\begin{equation} \label{quard_m} 
m^2  - 2 E^{'}m +({E^{'}}^2 - {C^{'}}^2) = 0.
\end{equation}
The solutions are 
\[
m = E^{'} \pm C^{'}.
\]

Restoring $m = z + \tfrac{1}{z}$, we obtain
\[
z+ \frac{1}{z} = E^{'} \pm C^{'},
\]
which yields the quadratic results
\begin{equation} \label{quard_z}
z^2 -(E^{'} \pm C^{'})z + 1 = 0.
\end{equation}
Solving this, we get:
\[
z = \frac{(E^{'} \pm C^{'}) \pm \sqrt{{(E^{'} \pm C^{'})}^2-4}}{2}.
\]

From this expression, we see that $z$ becomes complex when
\[
(E^{'} \pm C^{'})^2 - 4 < 0 \quad \Rightarrow \quad |E \pm C| < 2 t.
\]
Since $C = \lambda \cos(2 Q n^{\nu})$, so the value of $C$ is bounded within the interval $[-\lambda, \lambda]$. The localization criterion must hold for all possible values of $C$. Therefore, to ensure validity across the entire lattice, it is sufficient to enforce the inequality at the extremal values of C. So it is convenient to take the maximal value $C=\lambda$, which gives the final criterion:
\begin{equation} \label{cond_int}
| E \pm \lambda | < 2t,
\end{equation}
which corresponds to complex $z$ and hence extended (delocalized) wavefunctions \cite{sv_sds_1, MBL_Book}. Conversely, real $z$ arises when
\[
|E \pm \lambda| > 2t,
\]
indicating localized states. This exact analysis demonstrates the existence of a metal–insulator transition in the model in the presence of MEs, where the phase boundary of localized and delocalized states can be determined from Eq.~\ref{cond_int} as follows:
\begin{equation} \label{me_cond1}
-2t-\lambda < E  < 2t-\lambda ,
\end{equation}
and \begin{equation} \label{me_cond2}
-2t+\lambda < E  < 2t+\lambda ,
\end{equation}
with a critical point at $\lambda_c = 2t$. The rung modulation generates opposite effective onsite potentials in the two transformed channels, which underlie the emergence of mobility edges in the SVQL ladder. Introducing the symmetric and antisymmetric rung basis,
\[
\psi_n^\pm = \frac{1}{\sqrt{2}}(a_n \pm b_n),
\]
The ladder Hamiltonian decouples into two effective quasiperiodic channels with opposite onsite modulations. This gives us an alternative route to understand how mobility-edge physics can arise in a ladder geometry through bond-modulated inter-leg couplings, while both legs remain on-site homogeneous. Further details can be found in Appendix \ref{app_chan}.
\begin{figure*}[t]
    \centering
    \includegraphics[width=0.9\linewidth]{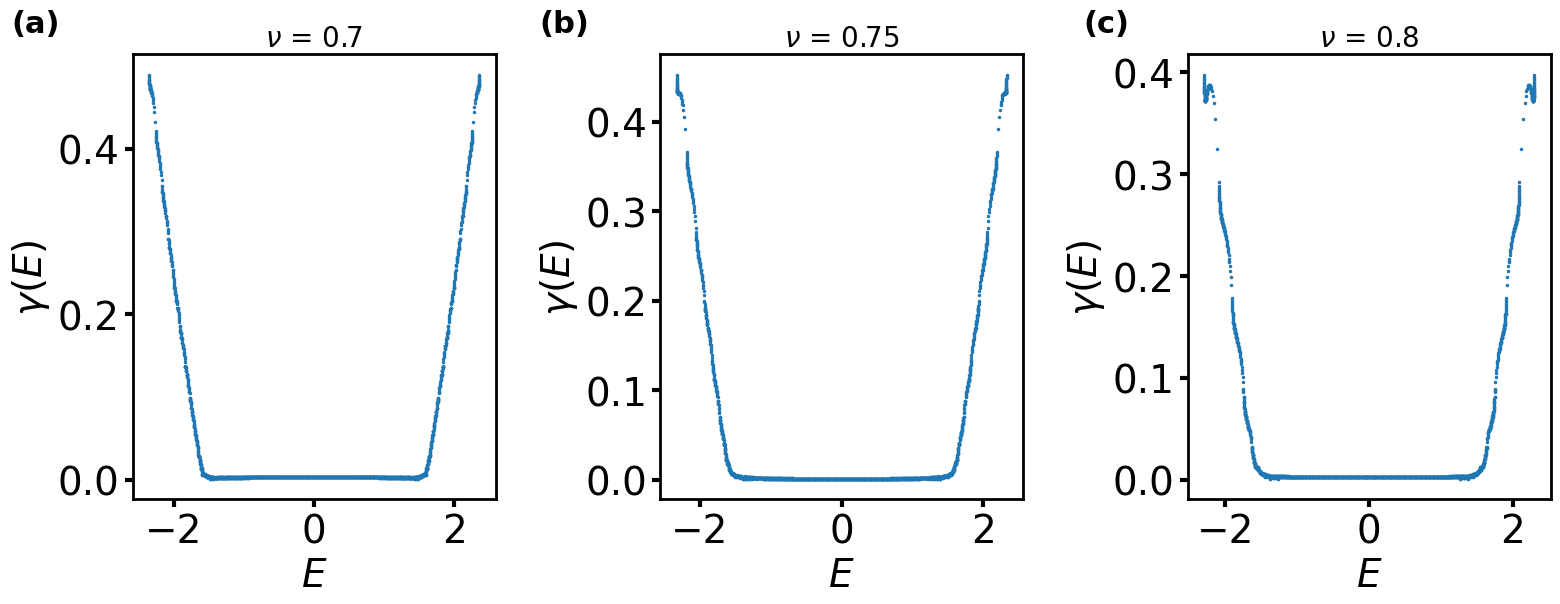}
    \caption{(a)-(c) are the plots of inverse localization length($\gamma$) as a function of energy eigenvalues ($E$) for $\nu=0.7, 0.75, 0.8$ respectively for $\lambda=0.4t$. Extended states correspond to $\gamma(E) \approx 0$, while localized states appear as finite $\gamma(E)$. Increasing $\nu$ enhances the sharpness of the ME, indicating stronger energy-selective localization in the SVQL model. For $\lambda=0.4t$, all the middle states are delocalized, and the ME appears at $E_c=-1.6t$, $+1.6t$, and $t$ is fixed to 1. We take $L=2000$ for numerics.}
    \label{lyapunov_0.4}
\end{figure*}
 
\begin{figure*}[t]
    \centering
    \includegraphics[width=0.95\linewidth]{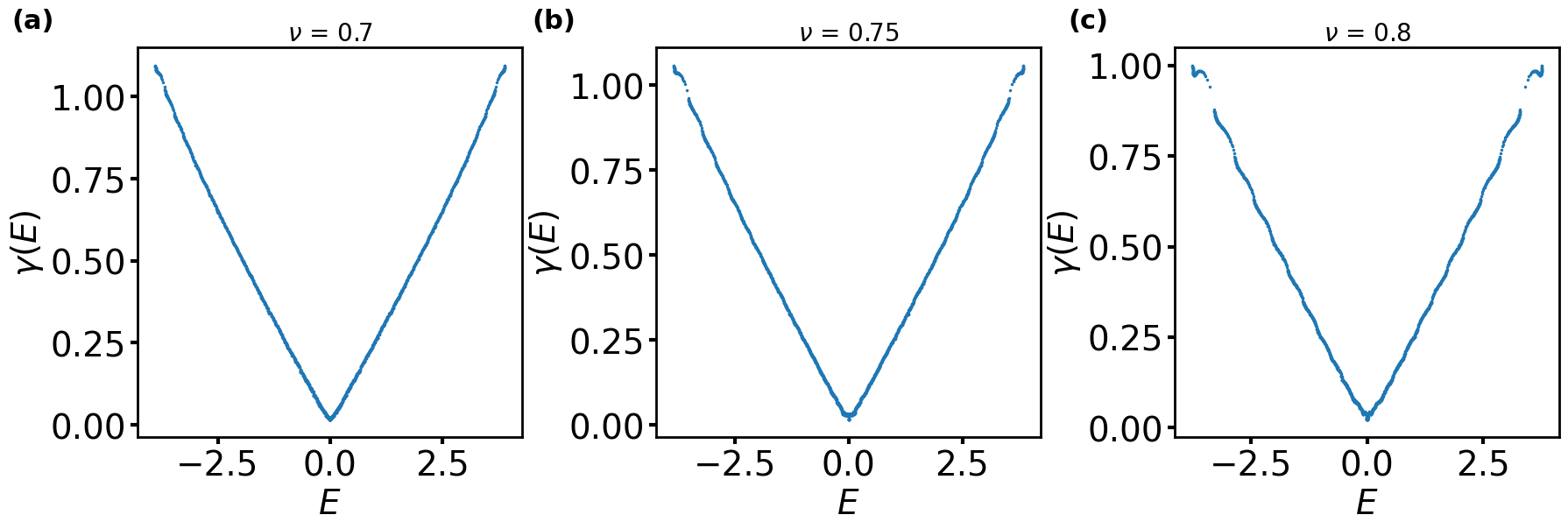}
    \caption{(a)-(c) represents the inverse localization length($\gamma$) as a function of eigen energy($E$) for $\nu=0.7,0.75, 0.8$ values respectively for $\lambda=2t$. In this case, the ME arises at $E_c=0,$ consistent with the derived expression $E_c=\pm|2t-\lambda|.$ We take $L=2000$ and $t=1$ for numerics.}
    \label{lyapunov_2}
\end{figure*}
\section{Numerical results}
\label{sec4}
\subsection{IPR: Inverse participation ratio}
To quantify the localization properties of the eigenstates, we evaluate the inverse participation ratio (IPR). For a normalized eigenstate of the Hamiltonian in Eq.~\ref{ham_eq},
\begin{equation}
|\psi_n\rangle = \sum_{i=1}^{L} \psi_n(i)\,|i\rangle, 
\qquad \sum_{i=1}^{L} |\psi_n(i)|^2 = 1,
\end{equation}
where $L$ denotes the total number of sites, the IPR is defined as
\begin{equation} \label{IPR_eq}
\mathrm{IPR} = 
\frac{\sum_{i=1}^{L} |\psi_n(i)|^{4}}
{\left(\sum_{i=1}^{L} |\psi_n(i)|^{2}\right)^2}.
\end{equation}

For extended states, $\mathrm{IPR} \sim 1/L \to 0$ in the thermodynamic limit, whereas for fully localized states $\mathrm{IPR} \to 1$. This limiting behavior can be illustrated by simple examples. In the case of complete localization at a single site, the probability amplitude satisfies $P_i = |\psi_n(i)|^2 = 1$ and $P_j = 0$ for all $j \neq i$, yielding $\mathrm{IPR} = 1$. For localization over two sites with equal probability, such as sites $a_i$ and $b_i$ on different legs, one has $P_{a_i} = P_{b_i} = 1/2$. Substituting into Eq.~\ref{IPR_eq} gives
\begin{equation}
\mathrm{IPR} = 
\frac{(1/2)^2 + (1/2)^2}{\left[(1/2) + (1/2)\right]^2} 
= \tfrac{1}{2}.
\end{equation}

The Fig.~\ref{fig:IPR_lambda}(a)–(c) present the IPR for all normalized eigenstates as a function of the bond-modulation strength $\lambda$ and the corresponding eigenenergies. On the other hand, Fig.~\ref{fig:IPR_lambda}(d)–(f) displays the IPR as a function of the normalized state index $k/L$, where $k$ is the eigenstate index. Both sets of plots reveal the presence of MEs at energies satisfying
\begin{equation}
E_c = \pm |2t -\lambda|.
\end{equation}
The analytically predicted MEs clearly separate localized from delocalized states. Notably, for $\lambda > 2t$, states near the band center begin to localize, signaling the onset of localization in the middle of the spectrum. While the bands of delocalized states get separated from the middle states by the MEs and follow the condition in Eq.~\ref{me_cond1} and Eq.~\ref{me_cond2}.

\begin{figure}[t]
    \centering
    \includegraphics[width=0.87\linewidth]{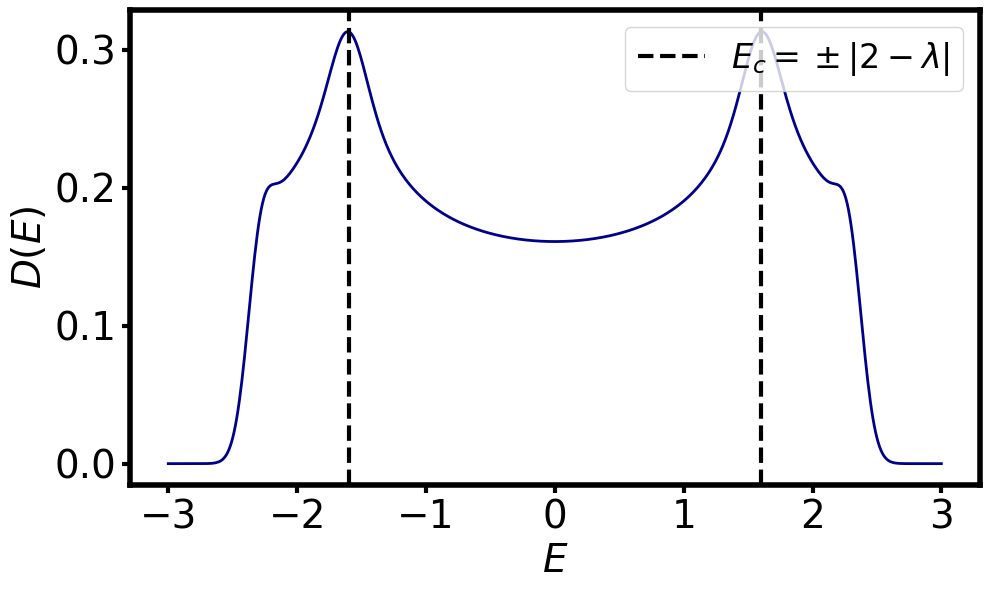}
    \caption{The average density of states (DOS) profile for $\nu=0.7$ at $\lambda=0.4t.$ The peaks in the DOS profile correspond to the MEs. The peaks are exactly located at the $E_c=\pm 1.6$, which exactly matches the analytical prediction. A system size of L = 2000 is used for the calculation, and we fixed $t$ to be 1. DOS computed via Gaussian-broadened spectra, averaged over 80 shifted-Q realizations (L=2000).}
    \label{dos_profile}
\end{figure}
\subsection{$\gamma(E)$: Inverse localization length}

We now turn to the characterization of localization properties using the inverse localization length $\gamma$, evaluated as a function of the eigenstate energy for different values of the modulation strength $\lambda$. The inverse localization length is computed according to \cite{sv_sds_1,sv_sds_2}:
\begin{equation}
    \gamma(E_{n}) = \frac{1}{L-1} \sum_{n' \ne n} 
    \ln \big|E_{n'} - E_{n}\big|,
    \label{gamma_eq}
\end{equation}
where $L$ denotes the total number of sites and $E_{n}$ is the eigenvalue corresponding to the $n^{th}$ eigenstate. This definition provides a convenient diagnostic: for extended states $\gamma(E) \to 0$ in the thermodynamic limit, whereas for localized states $\gamma(E)$ acquires a finite value. Thus, abrupt changes in $\gamma(E)$ as a function of energy signal the presence of MEs in the spectrum.

Figures~\ref{lyapunov_0.4}(a)–(c) display the variation of $\gamma$ with energy $E$ for modulation strengths $\nu = 0.7,~0.75,~0.8$ at fixed $\lambda = 0.4t$. In all cases, the inverse localization length remains nearly zero at the band center, consistent with extended states. As the energy approaches certain critical values $E_c$, however, $\gamma(E)$ exhibits a sharp jump from zero to a finite value, indicating the presence of MEs, and those particular $E_c$ values are identified as MEs. The plots clearly show that the MEs are at $E_c = \pm 1.6$ for these parameter sets. 

In contrast, Fig.~\ref{lyapunov_2} shows the behavior of $\gamma(E)$ for $\lambda = 2$. Here, regardless of the value of $\nu$, the ME shifts to the band center, with $E_c = 0$. This observation highlights the strong dependence of the localization transition on the interchain hopping strength. It is to be noted that for all numerical results we keep $t=1.$  

Importantly, the numerically extracted MEs are in excellent agreement with the analytical prediction obtained in Sec.~\ref{sec3}, namely
\begin{equation}
    E_c = \pm |2t - \lambda|.
\end{equation}
This consistency between analytical and numerical results confirms the robustness of the ME criterion and underscores the reliability of $\gamma(E)$ as a diagnostic for distinguishing localized and extended states across the spectrum.

\subsection{$D(E)$: Density of states} 
To further substantiate the presence of MEs, we examine the average density of states (DOS) $D(E)$ at critical energies ($E_c$). The DOS is defined as \cite{sv_sds_1}
\begin{equation}
    D(E) = \sum_{n=1}^{L} \delta(E - E_{n}),
    \label{dos_eq}
\end{equation}
where $E_{n}$ denotes the $n^{th}$ eigenenergy and $L$ is the total number of sites in the system. The DOS is directly related to the inverse localization length through the integral relation \cite{sv_sds_1}
\begin{equation}
    \gamma(E) = \int dE' \, D(E') \, \ln|E - E'|,
    \label{gamma_dos_eq}
\end{equation}
which provides a complementary perspective on the localization properties of the spectrum.

Figure~\ref{dos_profile} shows $D(E)$ as a function of $E$ for $\lambda = 0.4t$ and $\nu = 0.7$, with system size $L = 2000$. The DOS exhibits two pronounced peaks located precisely at the critical energies $E_c$, corresponding to the MEs identified in the previous sections. These peak-like singularities mark abrupt changes in the character of the eigenstates, signaling the transition from extended to localized behavior. Such sharp singular features are absent in the conventional three-dimensional Anderson localization problem \cite{anderson_3d, thouless}, where the average DOS is smooth over the $E_c$ values, underscoring the distinctive nature of the present model. The technical details of the average DOS calculation are presented in Appendix \ref{dos_cal}.

The correspondence between the singularities in $D(E)$ and the jumps observed in the inverse localization length $\gamma(E)$ further reinforces the interpretation of $E_c$ as genuine MEs. Thus, the extended–localized transition of the eigenstates can be consistently understood through the appearance of these sharp DOS singularities, in agreement with the analytical predictions and numerical diagnostics presented earlier. A scaling analysis of the DOS profile is presented in Appendix \ref{app_scal} to further strengthen the robustness argument. 
\begin{figure*}[t]
    \centering
    \includegraphics[width=1\linewidth]{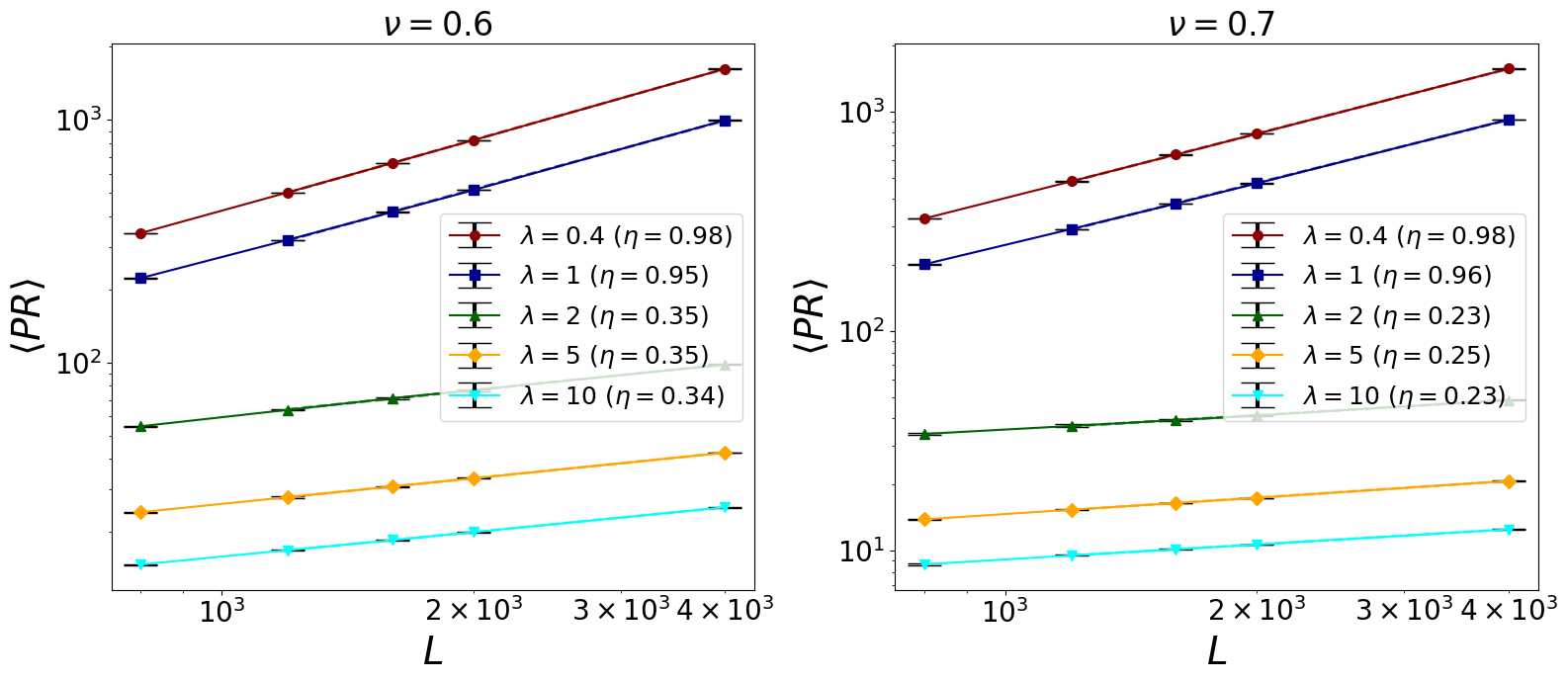}
    \caption{Log log plots for average participation ratio ($\langle PR \rangle$) as a function of system size (L). The inset shows the scaling exponent $\eta$ at different $\lambda$ values for $\nu=0.6$ and $0.7$. Increasing $\lambda$ reduces $D_2$, marking progressive localization, with sharper suppression for larger $\nu$, indicating stronger modulation-induced localization. At least 10 independent realizations have been performed in order to get better statistics, with error bars decreasing with system size.}
    \label{pr_l_small}
\end{figure*}
\begin{figure}[t]
    \centering
    \includegraphics[width=0.97\linewidth]{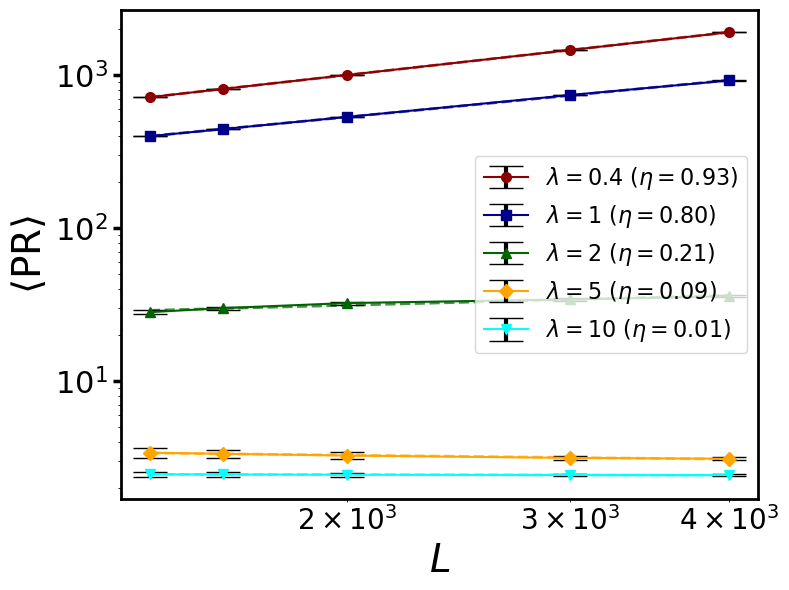}
    \caption{Log log plots for average participation ratio ($\langle PR \rangle$) as a function of system size (L). The inset shows the scaling exponent at different $\lambda$ values for $\nu=0.9$. At least 10 independent realizations have been performed in order to get better statistics, with error bars decreasing with increasing system size.}
    \label{pr_l_large}
\end{figure}
\subsection{$\langle PR \rangle$:Average participation ratio}

Finally, in this section, we elucidate the localization properties across different phases of the ladder system by performing a finite-size scaling analysis of the average participation ratio, $\langle PR \rangle$, defined as
\begin{equation} \label{pr_avr}
    \langle PR \rangle = \frac{1}{L} \sum_{n \in L} \frac{1}{IPR^{(n)}},
\end{equation}
where $IPR^{(n)}$ is the inverse participation ratio of the $n^{th}$ eigenstate as defined in Eq.~\ref{IPR_eq}, and the sum extends over all eigenstates of the Hamiltonian in Eq.~\ref{ham_eq}. For sufficiently large system sizes $L$, it is expected that $\langle PR \rangle$ obeys the scaling law
\begin{equation} \label{fsc_eta}
    \langle PR \rangle \sim L^{\eta},
\end{equation}
where $L$ denotes the number of sites and $\eta$ is the scaling exponent. This relation is particularly important for probing the multifractal nature of the eigenstates. More generally, the multifractal scaling is expressed as
\begin{equation} \label{fractal_dim}
    \frac{1}{L} \sum_{n \in L} 
    \left[\frac{1}{\sum_{j \in L} |\langle j|\psi_n\rangle|^{2q}}\right]
    \sim L^{D_q(q-1)},
\end{equation}
where $D_q$ is the fractal dimension associated with the $q-$th moment. The special case of Eq.~\ref{fsc_eta} corresponds to $q=2$, with $D_q = \eta$. The system is classified as delocalized when $\eta = 1$, localized when $\eta = 0$, and fractal when $0 < \eta < 1$ \cite{dynamics_2, tapan_reentrant2, Evers08}.

Figure~\ref{pr_l_small} illustrates the scaling of $\langle PR \rangle$ with system size $L$ for $\lambda = 0.4t, t, 2t, 5t,$ and $10t$. The corresponding log–log plots confirm that $\langle PR \rangle$ follows the scaling law of Eq.~\ref{fsc_eta} across all parameter regimes. Importantly, this scaling behavior remains valid even when the system simultaneously hosts delocalized (or weakly delocalized) and localized states, underscoring the robustness of the relation. Results are presented for three representative values of the modulation exponent $\nu$.

For small bond-modulation strength, $\lambda = 0.4t$ and $\lambda = t$, we obtain $\eta \approx 1$, confirming that the system resides in a delocalized phase. These delocalized states are robust against variations in $\nu$. At the critical point $\lambda = 2t$ (i.e., $\lambda = \lambda_c$), we find $0 < \eta < 1$, indicating a critical phase characterized by multifractal eigenstates. For $\lambda > \lambda_c$, the phase depends sensitively on the modulation exponent $\nu$. Specifically, for $\nu = 0.6$ and $0.7$, the system remains fractal, as the delocalized states near the band edges contribute to the growth of $\langle PR \rangle$ with system size. By contrast, for $\nu = 0.9$ and larger, the modulation is sufficiently strong to induce complete localization: $\eta \approx 0$, and $\langle PR \rangle$ ceases to grow with $L$. For all numerical results presented in this section, we set $t=1$. For $\nu=1$, all the states beyond $\lambda_c$ are completely localized.

These results demonstrate that finite-size scaling of $\langle PR \rangle$ provides a powerful diagnostic for distinguishing delocalized, localized, and fractal phases in the SVQL, and that the interplay between $\lambda$ and $\nu$ controls the nature of the eigenstates in a nontrivial manner. For small $\nu$ values, we confirmed the stability of the growth of average PR with system size up to $L=20000$. \par
Here, from the scaling analysis, we understand that the region just beyond the transition point $\lambda_c =2t$ is not a simple mixture of extended and localized states. This can be understood from the following argument: let us consider for a particular $\lambda$ value, the spectrum is a mixture of f fraction of localized states and (1-f) fraction of delocalized states. Then the average PR can be written as: \[\langle PR \rangle =f\langle PR\rangle_{L}+(1-f)\langle PR\rangle_{D}.\] Now, as the spectrum is just the mixture of localized and delocalized states, the scaling exponent will come from the delocalized component $\langle PR \rangle_{D}$, as the other component is of $O(1)$ ($\langle PR\rangle_{L} \sim O(1)$). So the overall $\langle PR \rangle$ will still exhibit a linear growth with system size (as $\langle PR \rangle _{D} \sim N$) as the largest scaling exponent dominates in the thermodynamic limit. Now, let us consider a situation in which we have a mixture of fractal and localized states. In this, the average PR will take the following form: \[\langle PR \rangle =f\langle PR\rangle_{L}+(1-f)\langle PR\rangle_{F}.\] In this case, also due to the above argument, the scaling will originate from the fractal component of the spectra ($\langle PR\rangle_{F} \sim N^{\eta}$, where $0<\eta<1$). Thus, according to the property of fractal states, the overall scaling will be $0<\eta <1.$ Hence, the system exhibits a critical phase in that region. 
\begin{figure*}[t]
    \centering
\includegraphics[width=1\linewidth]{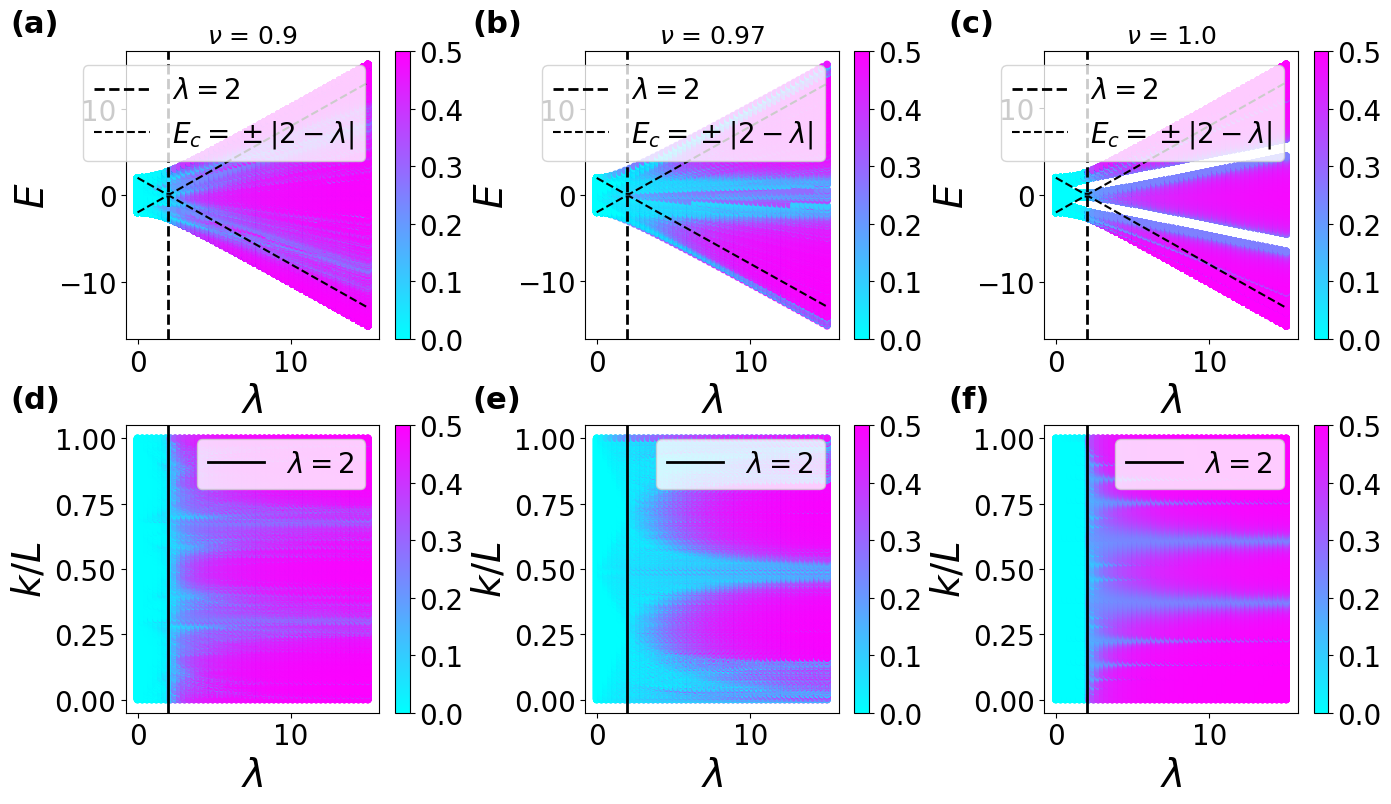}
    \caption{(a)-(c) are the contour plots of $IPR$ as function of $\lambda$ and energy ($E$) and (d)-(f) are the contour plots of $IPR$ as function of $\lambda$ and state index ratio($\frac{k}{L}$, where $k$ is the state index and $L$ is the total number of site) for all eigenstates. Calculations are performed for the total system size (L) = 2000 and for $\nu$ = 0.9, 0.97, and 1, respectively; $t$ is fixed to 1 for the calculations. The plots reveal a clear breakdown of the energy-dependent ME condition, obtained analytically at $\nu=0.97$, with low-IPR states still persisting in the middle of the spectrum.}
    \label{fig:IPR_lambda_rev1}
\end{figure*}
\section{Breakdown of locally constant approximation} \label{sec_rev_1}
\begin{figure}[t]
    \centering
\includegraphics[width=1\linewidth]{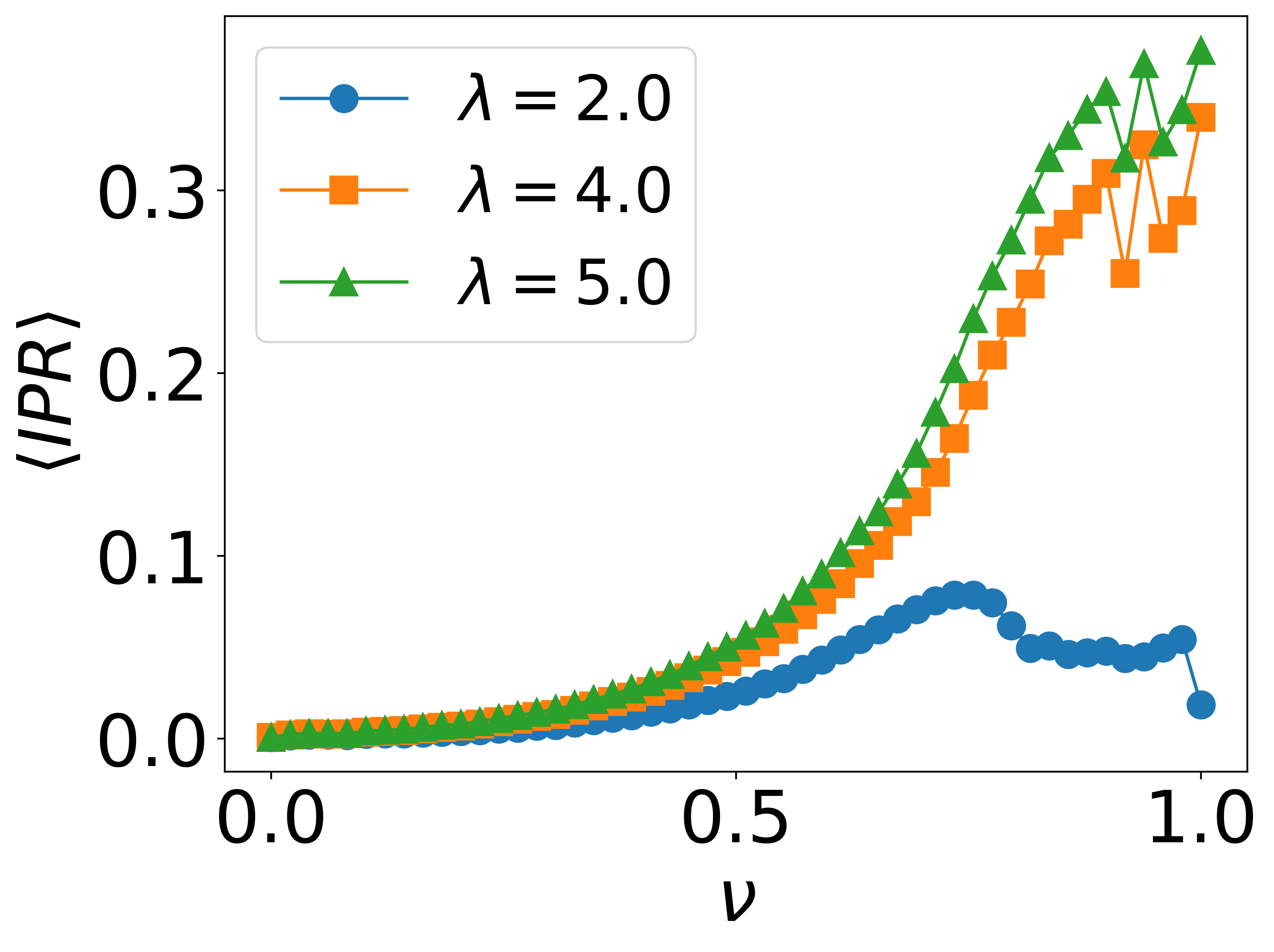}
    \caption{Plot of $\langle IPR \rangle$ as a function of $\nu$ for $\lambda=2t, 4t, 5t.$ Total system size $L$ is taken as $L=2000$ and $t=1$ for numerical calculation.}
    \label{fig:rev2}
\end{figure}

\begin{figure}[t]
    \centering
\includegraphics[width=1\linewidth]{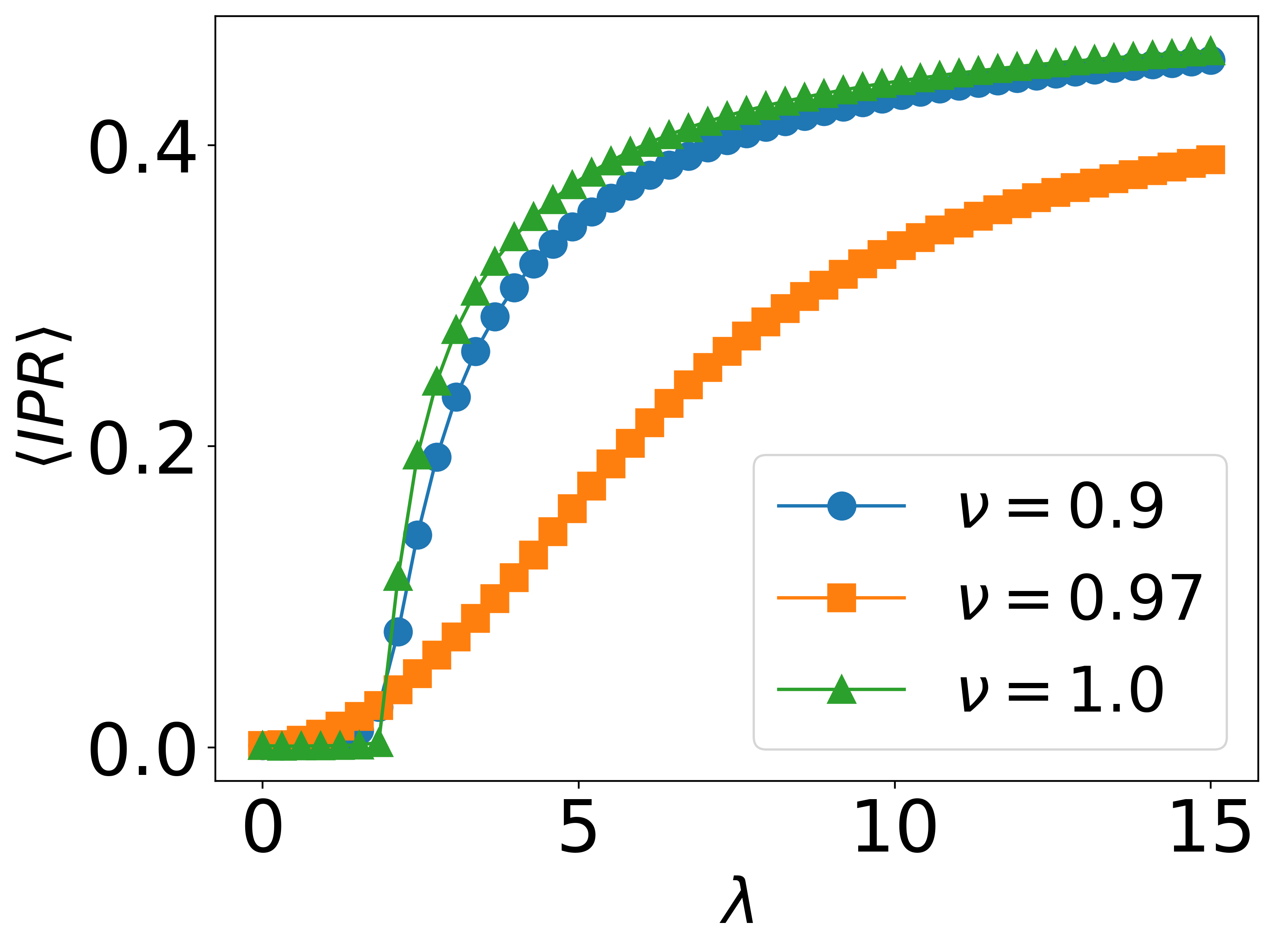}
    \caption{Plot of $\langle IPR \rangle$ as a function of $\lambda$ for $\nu=0.9,0.97, 1.$ Total system size is taken as $L=2000$ for numerical calculation.}
    \label{fig:rev3}
\end{figure}

The parameter $\nu$ controls the transition of the modulation from a periodic to a fully quasiperiodic regime, tuning the system from completely delocalized states to an AA-type localization. Our analytic ME condition relies on the locally constant approximation, which is valid for small $\nu$ and in the thermodynamic limit. It is therefore important to examine, in finite systems, the range of $0<\nu <1$ where this approximation breaks down.
We performed extensive numerical studies using several diagnostics to identify the crossover region in the $\nu$-parametric regime at large $\nu$ values. Figure~\ref{fig:IPR_lambda_rev1} shows the inverse participation ratio (IPR) of individual states as a function of $\lambda$ and energy (E) for $\nu=0.9,0.97, 1.0$, respectively. This deviation becomes most pronounced near $\nu=0.97$, where the onset of localization is significantly delayed compared to the analytical prediction. The results reveal deviations from the analytic ME condition: extended-like(low-IPR) states persist even beyond $\lambda =2t$ near the middle of the spectrum for $\nu=0.97$. The signature persists in both energy resolved IPR profile (see Fig.~\ref{fig:IPR_lambda_rev1}(a)-(c)) and state index resolved IPR plots (see Fig.~\ref{fig:IPR_lambda_rev1}(d)-(f)) as a function of $\lambda.$ We further compute the averaged IPR $\langle IPR \rangle$ across different $\nu$ values for three modulation strengths ($\lambda =2t,4t,5t$), shown in Fig.~\ref{fig:rev2}. For $\nu<1$, $\langle IPR\rangle$ increases monotonically with $\lambda$, consistent with the analytic prediction. However, near $\nu \rightarrow 1$, oscillations appear in $\langle IPR\rangle$, destabilizing the monotonic growth and indicating the persistence of low-IPR states beyond the critical point $\lambda =2t$.
Figure~\ref{fig:rev3} further shows $\langle IPR\rangle$ as a function of $\lambda$ for $\nu =0.9,0.97,1$. For fixed $\lambda$, the averaged IPR is consistently smaller at $\nu \approx 1$ than at $\nu =0.9$ and $1$, and this behavior persists even at large $\lambda$.\par 

This behavior identifies a crossover regime near $\nu \to1$ where the locally constant approximation breaks down in finite systems. In this regime, the ME position deviates from the analytical prediction, and extended-like states persist beyond the expected transition point. Importantly, this crossover does not correspond to a sharp phase transition but rather reflects a gradual interpolation between the slowly varying limit and the AA–like sharp transition. The effect is therefore intrinsic to finite system sizes and highlights the limitations of the asymptotic analytical description.

\begin{figure*}[t]
    \centering
\includegraphics[width=1\textwidth]{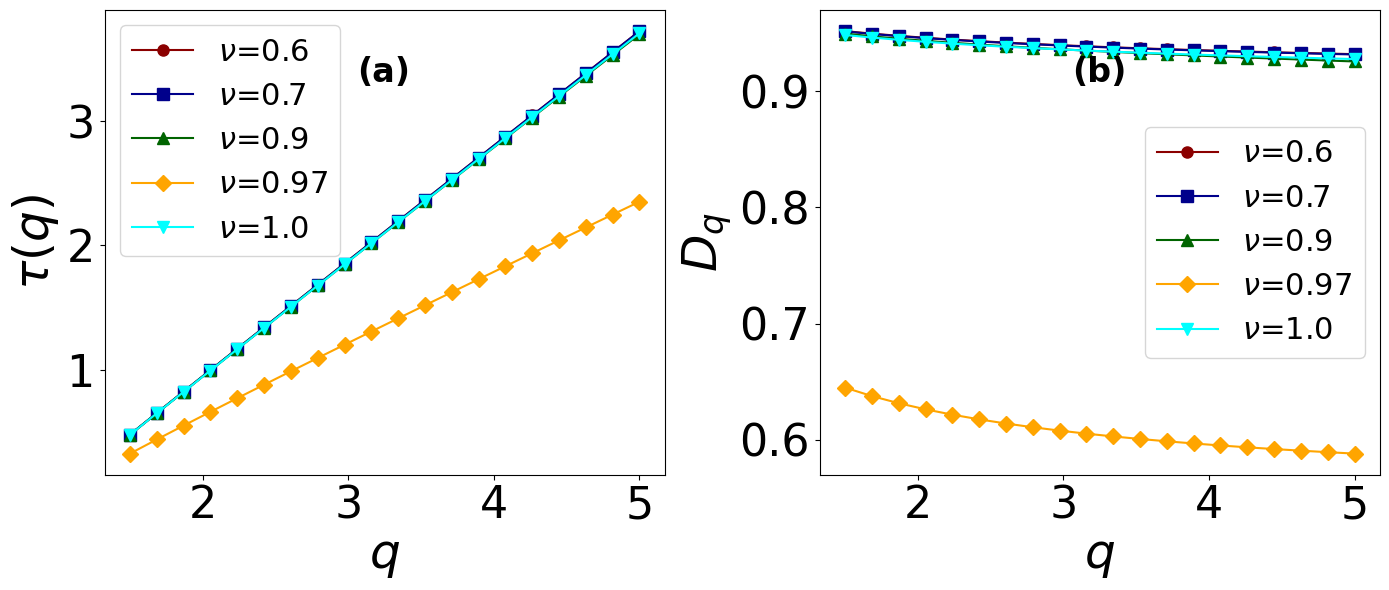}
    \caption{Plot of (a) $\tau(q)$ and (b) $D_q$ as a function of $q$ for $\nu=0.6, 0.7, 0.9, 0.97, 1$ at $\lambda=0.5t.$ Total system size is taken as $L=2000$ for numerical calculation.}
    \label{fig:rev4}
\end{figure*}
\begin{figure*}[t]
    \centering
\includegraphics[width=1\textwidth]{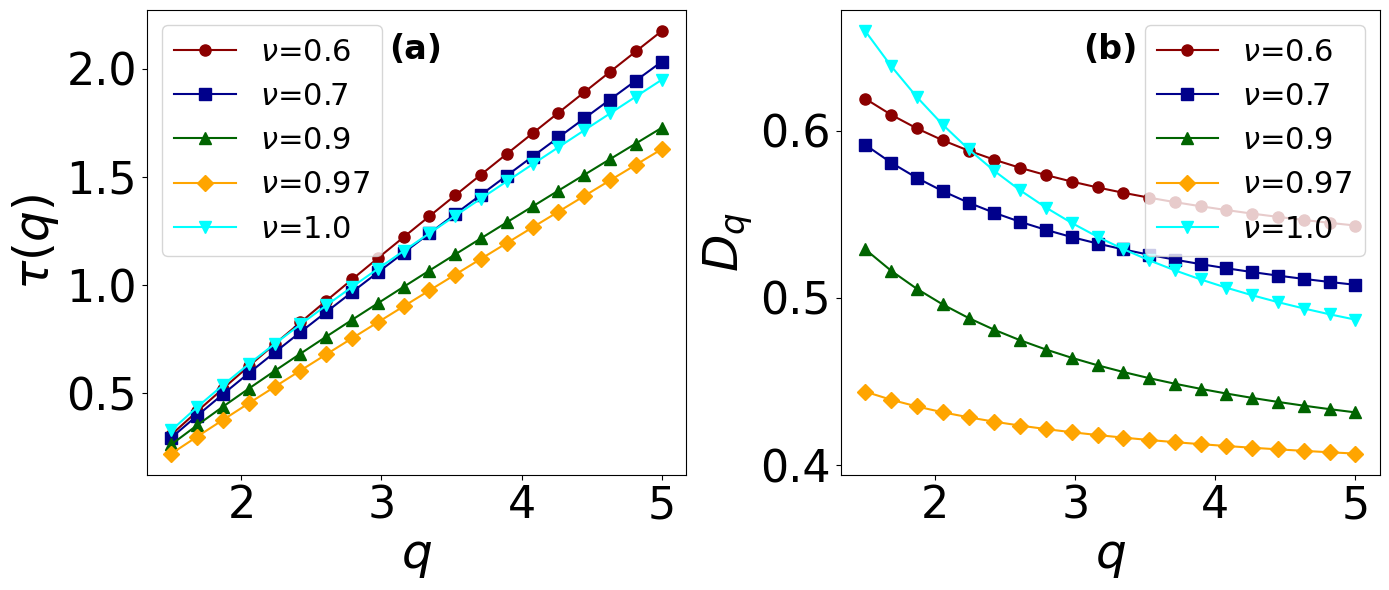}
    \caption{Plot of (a) $\tau(q)$ and (b) $D_q$ as a function of $q$ for $\nu=0.6, 0.7, 0.9, 0.97, 1$ at $\lambda=2t.$ Total system size is taken as $L=2000$ for numerical calculation.}
    \label{fig:rev5}
\end{figure*}
\begin{figure*}[t]
    \centering
\includegraphics[width=1\textwidth]{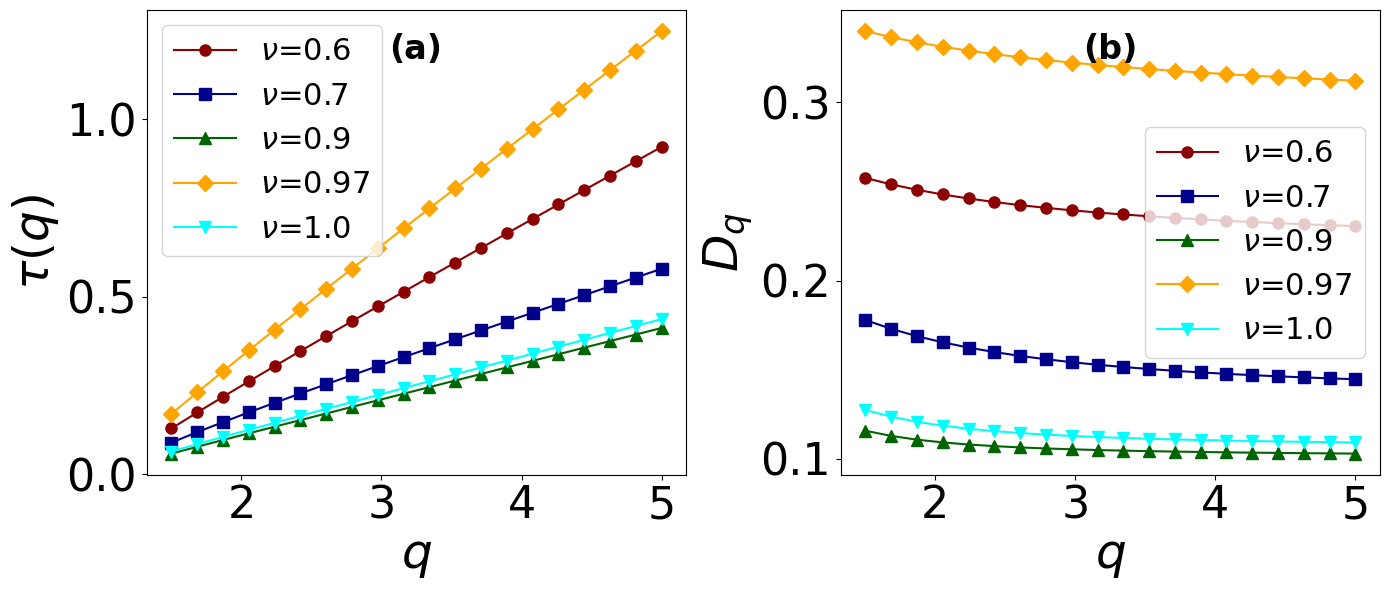}
    \caption{Plot of (a) $\tau(q)$ and (b) $D_q$ as a function of $q$ for $\nu=0.6, 0.7, 0.9, 0.97, 1$ at $\lambda=6t.$ Total system size is taken as $L=2000$ for numerical calculation.}
    \label{fig:rev6}
\end{figure*}

\section{Multifractal Analysis} \label{sec_rev_2}
To characterize the critical properties of the quasiperiodic ladder system, we perform a multifractal analysis of the eigenstates at three judiciously chosen $\lambda$ values (at $0.5t$, where the system is supposed to be in delocalized state, at $\lambda=2t$, an anlytically derived critical point and at $6t$, where the system is supposed to be in a strong localized region for large $\nu$, according to analytical prediction). This framework quantitatively distinguishes extended, localized, and critical states by examining how wavefunction amplitudes scale with system size. Two complementary quantities are introduced:
\begin{itemize}
    \item \textbf{Scaling exponent $\tau(q)$}: For a normalized eigenstate $\psi$, the generalized participation ratio is defined as   
\[
    P_q = \sum_i |\psi_i|^{2q}.
    \]
Its scaling with system size $N$ yields   
\[
    \tau(q) = -\lim_{N \to \infty}\frac{\ln P_q}{\ln N}.
    \]
    Extended states exhibit linear growth of $\tau(q)$ with $q$, localized states saturate, and critical states display nonlinear dependence, reflecting multifractality.
    \item \textbf{Fractal dimensions $D_q$}: From $\tau(q)$, we define the generalized fractal dimensions
\[
    D_q = \frac{\tau(q)}{q-1}.
    \]
    These dimensions quantify the effective support of the wavefunction. For extended states, $D_q \approx 1$; for localized states, $D_q \to 0$; and for multifractal states, $D_q$ assumes intermediate values that vary with $q$.
\end{itemize}
Figure~\ref{fig:rev4}(a) shows that for all $\nu$ values except $\nu=0.97$, $\tau(q)$ grows linearly, indicating delocalization for $\lambda=0.5t$. At $\nu=0.97$, however, the growth deviates from linearity. This is confirmed in Fig.~\ref{fig:rev4}(b): for $\nu=0.6, 0.7, 0.9, 1$, $D_q \simeq 1$ (for all $q$ values), consistent with extended states, while for $\nu=0.97$, $0 < D_q < 1$, revealing fractal character.
Figures~\ref{fig:rev5}(a) and \ref{fig:rev5}(b) demonstrate clear multifractal signatures at $\lambda=2t$. For smaller quasiperiodic exponents ($\nu=0.6, 0.7$), $\tau(q)$ grows strongly with $q$ where the linear dependence deviates significantly from that observed at $\lambda=0.5t$, and $D_q$ remains relatively high (though strictly less than $1$), indicating fractality. As $\nu$ increases ($\nu=0.9$), the scaling becomes more non-linear (compared to $\nu=0.6, 0.7$) and $D_q$ decreases, signaling residual fractality. Near $\nu=0.97$, both $\tau(q)$ and $D_q$ drop sharply, with $D_q > 0$ but remains very small. Finally, at $\nu=1.0$ (the AA limit), the states exhibit $0 < D_q < 1$ decreasing with $q$, consistent with the known critical multifractal nature of eigenstates at the AA critical point $\lambda_c=2t$. 
Figure~\ref{fig:rev6}(a) shows that for $\nu=0.6, 0.7, 0.9, 1$, $\tau(q)$ growth is strongly suppressed, indicating localization at $\lambda=6t$. Remarkably, $\nu=0.97$ retains a strong non-linear dependence of $\tau(q)$ with comparatively faster growth. Figure~\ref{fig:rev6}(b) complements this: for $\nu=0.6, 0.7$, $D_q$ is very small; for $\nu=0.9, 1$, $D_q \simeq 0$, consistent with the complete localization; but at $\nu=0.97$, $D_q$ remains comparatively higher, suggesting persistence of fractal character. 
Thus, $\nu=0.97$ should be interpreted as a crossover regime, where the nature of multifractal scaling changes most rapidly, interpolating between the slowly varying limit and the AA limit. We emphasize that this crossover is not indicative of a sharp phase transition but rather reflects the gradual breakdown of the locally constant approximation. A systematic finite-size scaling analysis would be required to establish the asymptotic behavior of the multifractal exponents in this regime.

\section{Conclusion}
\label{cons}
We have introduced and analyzed a minimal two-leg slowly varying quasiperiodic ladder system in which the ladder realizes mobility-edge physics through deterministic modulation of the inter-leg couplings. In contrast to conventional quasiperiodic ladder models with explicit onsite modulation, the present construction realizes mobility-edge physics through bond-modulated ladder couplings while both legs remain onsite homogeneous. Within the slowly varying regime ($0 < \nu < 1$), the system admits an analytical treatment that yields the exact mobility-edge condition
\[E_c=\pm|2t-\lambda|.\]
Beyond the analytical mobility-edge condition, we identified the finite-size crossover associated with the breakdown of the locally constant approximation and characterized the multifractal properties of the eigenstates across the spectrum. Combined with extensive numerical calculations, our results establish the SVQL model as a minimal and analytically tractable ladder platform for studying quasiperiodic mobility-edge physics. The model demonstrates that mobility edges can arise solely from slowly varying bond modulation, without requiring explicit quasiperiodic onsite potentials in the underlying microscopic Hamiltonian.

Moreover, such ladder geometries are commonly realized in cold atom optical lattices, photonic waveguide arrays, and engineered solid-state networks.  Slowly varying inter-leg tunneling can be engineered in cold-atom ladder systems using long-wavelength superlattice modulations \cite{Kang_2020}, Raman-assisted tunneling with spatially graded phases \cite{Li2022}, or Floquet protocols. Photonic waveguide arrays~\cite{expt_1} offer an equally promising platform, where the spacing between waveguides can be varied smoothly to produce the required bond modulation. The coexistence of extended and localized modes, tunable via $\lambda$, suggests that such systems may serve as controlled settings for probing nonergodic extended phases, subdiffusive transport, and quasiperiodic analogs of ME-assisted localization. This opens the path for directly testing the predictions and even the stability of these MEs in the presence of interactions.
\par 
In cold-atom~\cite{EXP_1,EXP_2} ladder systems, the coexistence of localized and extended states can be probed through real-space expansion dynamics or momentum-resolved imaging, enabling direct experimental access to the predicted mobility edge condition: $E_c=\pm |2t-\lambda |$. In particular, the mobility edge can be identified by preparing wave packets with controlled energy distributions and monitoring their long-time spreading behavior across the predicted threshold. Photonic waveguide arrays provide a complementary platform in which light propagation directly visualizes the spatial structure of eigenmodes, allowing simultaneous observation of localized and extended optical states as a function of the modulation strength. In engineered solid-state architectures, such as superconducting qubit arrays~\cite{sc_qubit_1, sc_qubit_2}, tunable inter-site couplings can be used to implement rung-modulated ladder geometries, while spectroscopic measurements can be used to extract participation ratios across the spectrum. Together, these platforms provide concrete, experimentally feasible pathways to test our theoretical predictions and investigate the robustness of mobility edges in the presence of interactions and finite-size effects. We also note that the present bond-modulated ladder geometry may offer practical advantages for experimental realization. In several platforms, including optical lattices, photonic waveguide arrays, and superconducting-circuit architectures, coupling strengths can often be engineered and tuned in a controlled manner through lattice geometry, waveguide spacing, or circuit design. By contrast, implementing spatially varying quasiperiodic onsite potentials typically requires additional superlattice engineering or site-selective parameter control. Consequently, the ladder construction studied here provides a potentially accessible route for exploring mobility-edge physics arising from quasiperiodic bond modulation.

Our numerical analysis—combining inverse participation ratios, Lyapunov exponents, and density-of-states calculations—shows excellent agreement with the analytical prediction. Across a wide range of parameters, the system exhibits a mixed phase in
which localized and extended eigenstates coexist. The sharpness of the ME improves significantly for smaller $\nu$, consistent with the increasing accuracy of
the locally constant approximation. As $\nu\rightarrow 1$, the model continuously connects to a rung-modulated Aubry--André ladder and exhibits a single critical point
at $\lambda_c=2t$, thereby recovering the expected AAH-like behavior.

Several open directions naturally follow from our work. First of all, the ladder geometry
invites exploration of many-body localization in the presence of inter-leg quasiperiodicity, where the two-channel structure may influence Fock-space connectivity in nontrivial ways. The role of interactions in selectively hybridizing symmetric and antisymmetric channels also remains largely unexplored. Even the extensions to multi-leg ladders or two-dimensional strip geometries may exhibit new ME structures arising from higher-order band interference. Finally, connections to non-Hermitian quasiperiodicity, topological MEs, and transport in the nonergodic metallic regime offer promising directions for further exploration.

Overall, our results demonstrate that slowly varying quasiperiodicity in inter-leg couplings provides a robust and analytically tractable route to mobility-edge physics in ladder systems. Beyond establishing an exact mobility-edge condition and detailed characterization of the associated phases, the ladder geometry provides a distinct microscopic realization of slowly varying quasiperiodic mobility-edge physics. Through the symmetric–antisymmetric rung transformation, the model maps onto two effective, slowly varying quasiperiodic chains, enabling an analytical description of the mobility edge while preserving a physically relevant ladder implementation. We expect this framework to serve as a useful platform for future investigations of interacting, non-Hermitian, and higher-dimensional quasiperiodic systems, as well as for experimental studies in cold-atom, photonic, and engineered quantum-lattice settings.

\noindent
\section{Acknowledgments}
AG gratefully acknowledges Dr. Shaon Sahoo for his valuable suggestions and support throughout this research. The author also thanks Dr. Ranjan Modak for insightful discussions and helpful suggestions. AG is also greatly thankful to the research facilities provided by IIT Tirupati. 

\appendix
\begin{figure}[t]
    \centering
\includegraphics[width=0.5\textwidth]{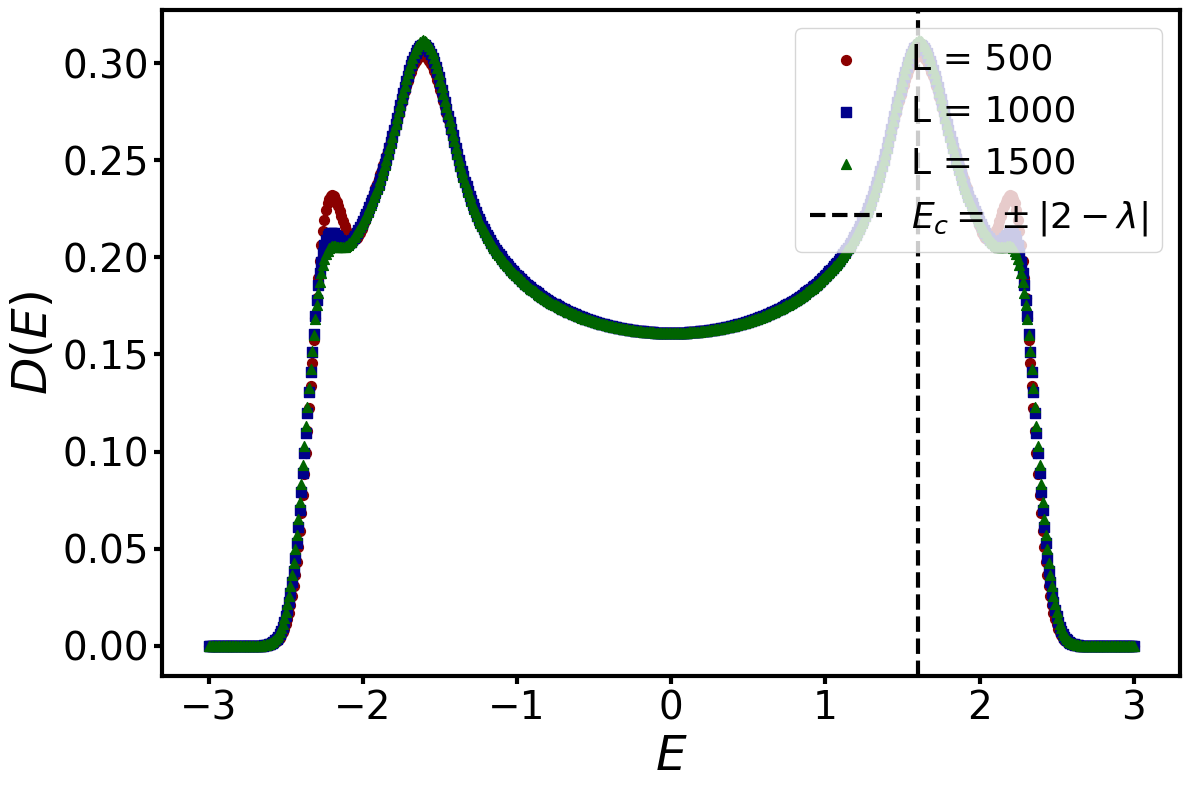}
    \caption{System-size scaling of average DOS profile with system size $L=500, 1000, 1500,$ for $\nu=0.7$ at $\lambda=0.4t,$ with $t=1$.}
    \label{dos_scal}
\end{figure}
\section{Symmetric--Antisymmetric Channel Decomposition and Effective Hamiltonian \label{app_chan}} 

In this Appendix, we explicitly demonstrate how the two-leg ladder with
slowly varying quasiperiodic rung modulation decomposes into two
independent one-dimensional channels. This reveals the microscopic
origin of the $\pm \Gamma_n$ effective onsite potentials that underlie the ME condition derived in Sec.~III.

\subsection{Transformation to symmetric and antisymmetric basis}

We define the standard rung-mode transformation,
\begin{equation}
\psi_n^{\pm} = \frac{1}{\sqrt{2}} \left( a_n \pm b_n \right),
\label{eq:S1}
\end{equation}
with inverse relations
\begin{equation}
a_n = \frac{\psi_n^{+} + \psi_n^{-}}{\sqrt{2}}, 
\qquad
b_n = \frac{\psi_n^{+} - \psi_n^{-}}{\sqrt{2}}.
\label{eq:S2}
\end{equation}
Here $\psi_n^{+}$ denotes the symmetric channel and $\psi_n^{-}$ the
antisymmetric channel.

\subsection{ Ladder Hamiltonian in matrix form}

The single-particle Schr\"odinger equation for the amplitudes
$(a_n,b_n)$ reads:
\begin{equation}
E 
\begin{pmatrix} 
a_n \\[2pt] b_n 
\end{pmatrix}
=
t
\begin{pmatrix}
a_{n+1}+a_{n-1} \\[2pt]
b_{n+1}+b_{n-1}
\end{pmatrix}
+
\Gamma_n
\begin{pmatrix}
0 & 1 \\
1 & 0
\end{pmatrix}
\begin{pmatrix}
a_n \\[2pt] b_n
\end{pmatrix},
\label{eq:S3}
\end{equation}
where $\Gamma_n = \lambda \cos(2 Q n^\nu)$ is the slowly varying
quasiperiodic rung coupling.

\subsection{Diagonalization of the rung subspace}

Define the unitary transformation
\[
U=\frac{1}{\sqrt{2}}
\begin{pmatrix}
1 & 1\\[2pt]
1 & -1
\end{pmatrix},\qquad
\begin{pmatrix}\psi_n^{+}\\[2pt]\psi_n^{-}\end{pmatrix}
=U\begin{pmatrix}a_n\\[2pt]b_n\end{pmatrix}.
\]
In the original rung basis $(a_n,b_n)$, the inter-leg coupling is
\[
T=\begin{pmatrix}0&1\\[2pt]1&0\end{pmatrix}.
\]
Transforming to the symmetric/antisymmetric basis yields
\[
U^\dagger T U=\begin{pmatrix}1&0\\[2pt]0&-1\end{pmatrix},
\]
so the symmetric channel has eigenvalue $+1$ and the antisymmetric channel has eigenvalue $-1$. Including the physical amplitude $\Gamma_n$, one has
\[
U^\dagger (\Gamma_n T) U=\begin{pmatrix}\Gamma_n&0\\[2pt]0&-\Gamma_n\end{pmatrix}.
\]
\subsection{Decoupled effective chain equations}

Transforming Eq.~(\ref{eq:S3}) using Eqs.~(\ref{eq:S1})--(\ref{eq:S2}),
we obtain two decoupled equations:
\begin{align}
E \psi_n^{+} &= 
t (\psi_{n+1}^{+} + \psi_{n-1}^{+}) 
+ \Gamma_n \psi_n^{+},
\label{eq:S4}
\\[4pt]
E \psi_n^{-} &= 
t (\psi_{n+1}^{-} + \psi_{n-1}^{-}) 
- \Gamma_n \psi_n^{-}.
\label{eq:S5}
\end{align}
Thus, the ladder maps to two independent chains, each experiencing
opposite effective onsite potentials:
\begin{equation}
V_n^{+} = +\Gamma_n, 
\qquad 
V_n^{-} = -\Gamma_n.
\end{equation}

\subsection{ Slowly varying limit and local constancy}

For $0 < \nu < 1$, the modulation satisfies
\begin{equation}
\Gamma_{n+1} - \Gamma_n \rightarrow 0 
\qquad (n \rightarrow \infty),
\end{equation}
so the modulation may be treated as \emph{locally constant}:
\begin{equation}
\Gamma_n \approx C,
\qquad C \in [-\lambda,\lambda].
\label{eq:S6}
\end{equation}
Using Eq.~(\ref{eq:S6}), Eqs.~(\ref{eq:S4})--(\ref{eq:S5}) reduce to
\begin{align}
(E - C)\psi_n^{+} &= t(\psi_{n+1}^{+}+\psi_{n-1}^{+}),
\label{eq:S7}
\\
(E + C)\psi_n^{-} &= t(\psi_{n+1}^{-}+\psi_{n-1}^{-}).
\label{eq:S8}
\end{align}
Each channel, therefore, behaves as a shifted one-dimensional tight-binding
model with bandwidth $4t$.

\subsection{ME condition}

Extended states require the effective energies to lie inside the band:
\[
|E - C| < 2t,
\qquad
|E + C| < 2t.
\]
Maximizing over all $C \in [-\lambda,\lambda]$ yields
\begin{equation}
|E \pm \lambda| = 2t.
\end{equation}
So the exact ME positions are
\begin{equation}
\boxed{
E_c = \pm |2t - \lambda|.
}
\label{eq:MEfinal}      
\end{equation}
\\
\subsection{Physical interpretation}
Thus, the ladder Hamiltonian can be diagonalized into symmetric and antisymmetric channels, yielding two effective one-dimensional equations with opposite quasiperiodic onsite potentials studied in Ref.~\cite{sv_sds_1}. It operates at the single-particle level. The present construction demonstrates that the ladder system provides a microscopic realization of exact mobility-edge physics with quasiperiodicity introduced solely through inter-leg tunneling, rather than through explicit on-site modulation. In this sense, the SVQL model provides an alternative microscopic implementation of mobility-edge physics in a multi-channel geometry while remaining analytically tractable in the slowly-varying limit.

\section{Density of States Calculation\label{dos_cal}}
The density of states (DOS) was obtained via exact diagonalization of the Hamiltonian for a given system size $L$, where $Q$ is a quasiperiodic phase parameter varied across a realization of $N_{\text{real}}=80$, to ensure statistical averaging. For each realization, the eigenvalues $\{E_{n}\}$ were computed using exact diagonalization. The DOS was 
constructed on a fixed energy grid by Gaussian broadening of each eigenvalue,
\[
D(E) = \frac{1}{N_{\text{real}} \cdot L} 
\sum_{r=1}^{N_{\text{real}}} \sum_{n=1}^{L} 
\frac{1}{\sqrt{2\pi}\,\sigma} 
\exp\!\left[-\frac{(E - E_{n}^{(r)})^{2}}{2\sigma^{2}}\right],
\]
with a broadening width $\sigma = 0.1$, and $E_{n}^{r}$ is the eigen value of the $n^{th}$ eigenstate at $r^{th}$ realization. The averaged DOS was normalized by the total number of states and 
realizations. Analytical mobility edge estimates were indicated at
\[
E_{c} = \pm |2t - \lambda|.
\]
\section{Scaling of DOS profile with system size(L) \label{app_scal}}
To assess the robustness of the density of states (DOS) against finite-size effects, we computed the DOS for system sizes L=500, 1000, and 1500 in fig \ref{dos_scal}. The results demonstrate a clear data collapse in the DOS profile for different system sizes. We observe that the peak positions in the DOS profile do not shift with varying system size, indicating that the extracted ME position is stable against the system sizes considered.
\nocite{*}

\bibliography{manuscript}

\end{document}